\documentclass[10pt,a4paper]{article}
\usepackage{graphicx, array, blindtext}
\usepackage[T1]{fontenc}
\usepackage[utf8]{inputenc}
\usepackage{authblk}
\usepackage{anyfontsize}
\usepackage{epstopdf}
\usepackage{authblk}
\usepackage{anyfontsize}
\usepackage{authblk}
\usepackage{amssymb,amsmath}
\usepackage{amsmath}

\title{Interacting varying Ghost Dark energy models in General Relativity}
\author[1]{Martiros Khurshudyan\thanks{khurshudyan@yandex.ru, martiros.khurshudyan@mpikg.mpg.de}}
\author[2]{Amalya Khurshudyan\thanks{amalya.khurshudyan@uni.lu}}

\affil[1]{\footnotesize{Max Planck Institute of Colloids and Interfaces,\\ Potsdam-Golm Science Park Am Muhlenberg 1 OT Golm, 14476
Potsdam}}
\affil[3]{\footnotesize{University of Luxembourg, Faculte des Sciences, de la Technologie et de la Communication, Luxembourg}}

\begin{document}
  \maketitle

\begin{abstract}
Motivated by recent developments in Cosmology we would like to consider an extension of the Ghost DE which we will call as varying Ghost DE. Ghost DE like other models was introduced recently as a possible way to explain accelerated expansion of the Universe. For the phenomenological origin of the varying Ghost dark energy in our Universe we can suggest an existence of some unknown dynamics between the Ghost Dark energy and a fluid which evaporated completely making sense of the proposed effect. Moreover, we assume that this was in the epochs and scales which are unreachable by nowadays experiments, like in very early Universe. In this study we will investigate the model for cosmological validity. We will apply observational
and causality constraints to illuminate physically correct behavior of the model from the phenomenological one. We saw that an interaction between the varying Ghost DE and cold DM~(CDM) also provides a solution to the cosmological coincidence problem. And we found that the Ghost DE behaves as a matter like fluid in early Universe.
\end{abstract}

\section{\large{Introduction}}
The observations of high redshift type SNIa supernovae~\cite{Riess}~-~\cite{Riess2} reveal the speeding up expansion of our Universe. The surveys of clusters of galaxies showed that the density of matter is very much less than critical density~\cite{Pope}, observations of Cosmic Microwave Background anisotropies indicate that the Universe is flat and the total energy density is very close to the critical $\Omega_{\small{tot}} \simeq1$~\cite{Spergel}. In order to explain experimental data concerning to the nature of the accelerated expansion of the Universe several hypothesis were proposed. For instance, in General Relativity framework, the desirable result could be achieved by dark energy: an exotic and mysterious component of the Universe, with negative pressure and negative EoS parameter $\omega<0$. Dark energy occupies about 73$\% $ of the energy of our Universe, other component, Dark matter, about 23$\%$, and usual baryonic matter occupies about 4$\%$. The simplest model for the dark energy is the cosmological constant $\omega_{\Lambda}=-1$ introduced by Einstein, but  with absence of a fundamental mechanism which sets the cosmological constant zero or very small value. This problem known as fine-tuning problem, because in the framework of quantum field theory, the expectation value of the vacuum energy is 123 order of magnitude larger than the observed value of the $\Lambda$~\cite{Steinhardt}. The second problem known as cosmological coincidence problem, which asks why are we living in an epoch in which the densities of dark energy and matter are comparable? Alternative models of dark energy suggest a dynamical form of dark energy, which at least in an effective level, can originate from a variable cosmological constant~\cite{Sola}, or from various fields~\cite{Ratra}~-~\cite{Ratra6},~\cite{Caldwell}~-~\cite{Caldwell9},~\cite{Feng}~-~\cite{Feng13} and could alleviate these problems. Finally, an interesting attempt to probe the nature of dark energy according to some basic quantum gravitational principles are the holographic dark energy paradigm~\cite{Hsu}~-~\cite{Hsu11} and agegraphic dark energy models~\cite{Cai1}~-~\cite{Cai13}.
New model of dark energy called Veneziano ghost dark energy has been recently proposed, which supposed to solve the $U(1)_{A}$ problem in low-energy effective theory of QCD~\cite{Ghost1}~-~\cite{Chao-Jun2}. Indeed, the contribution of the ghosts field to the vacuum energy in curved space or time-dependent background can be regarded as a possible candidate for the dark energy. Veneziano ghost exhibits non trivial physical effects in the expanding Universe and these effects give rise to a vacuum energy density $\rho_{D}\sim \Lambda^{3}_{QCD}H\sim (10^{-3}eV)^{4}$. With $H\sim 10^{-33}eV$ and $\Lambda_{QCD}\sim 100 eV$ we have the right value for the force accelerating the Universe today. Energy density of the Ghost Dark energy reads as
\begin{equation}\label{eq:GDE}
\rho_{\small{G}}=\alpha H,
\end{equation}
where $H$ is Hubble parameter $H=\dot{a}/a$ and $\alpha$ is a constant parameter of the model, which should be determined. A generalization of the model~\cite{Cai} also was proposed for which energy density reads as
\begin{equation}\label{eq:GDEgen}
\rho_{\small{G}}=\alpha H+\beta H^{2},
\end{equation}
with $\alpha$ and $\beta$ constant parameters of the model. In this work we would like to suggest a phenomenological modification of the Ghost DE and investigate cosmological consequences of such modification. Let suppose that our Universe with FRW metric contains a mixture of a varying Ghost DE and a barotropic fluid with
\begin{equation}
P_{m}=\omega\rho_{m},
\end{equation}
EoS. We know that the energy conservation for the composed fluid reads as
\begin{equation}\label{eq:energy}
\dot{\rho}+3H(\rho+P)=0,
\end{equation}
with total energy density and pressure of composed fluid given as
\begin{equation}\label{eq:fluid}
\rho=\rho_{\small{VG}}+\rho_{m} ~~~~~~{\rm and}~~~~~~ P=P_{\small{VG}}+P_{m}.
\end{equation}
Field equations with FRW metric read as
\begin{equation}\label{eq: Fridmman vlambda}
H^{2}=\frac{\dot{a}^{2}}{a^{2}}=\frac{8\pi G\rho}{3},
\end{equation}
\begin{equation}\label{eq:fridman2}
\frac{\ddot{a}}{a}=-\frac{4\pi G}{3}(\rho+3P).
\end{equation}
We suppose, that the cosmological constant $\Lambda=0$, the gravitational constant $G$ and $c$ are constants with $c=8\pi G=1$. In modern cosmology an interaction between fluid components plays an important role and mathematically means
\begin{equation}\label{eq:firstfluid}
\dot{\rho}_{m}+3H(\rho_{m}+P_{m})=Q
\end{equation}
and
\begin{equation}\label{eq:secondfluid}
\dot{\rho}_{\small{G}}+3H(\rho_{\small{G}}+P_{\small{G}})=-Q,
\end{equation}
where $Q$ denotes the phenomenological interaction. From the thermodynamical point of view, it is argued that the second law of thermodynamics strongly faviours dark energy decaying into dark matter. Usually, considered  forms for the interaction term $Q$ are
\begin{equation}\label{eq:Q1}
Q=3Hb\rho_{d},
\end{equation}
\begin{equation}\label{eq:Q2}
Q=3Hb(\rho_{d}+\rho_{m}),
\end{equation}
and
\begin{equation}\label{eq:Q2}
Q=3Hb\rho_{m},
\end{equation}
where $b>0$ is a coupling constant. Mentioned interaction terms are either positive or negative and can not change the sign during the evolution of the Universe. However, recently by using a model independent method to deal with the observational data was found that the sign of the interaction term $Q$ in the dark sector changed in the redshift range of $0.45 \lesssim z \lesssim 0.9$.  Hereafter, a sign-changeable interaction~\cite{Hao}~-~\cite{Hao2}, were introduced
\begin{equation}\label{eq:signcinteraction}
Q=q(\alpha\dot{\rho}+3\beta H\rho).
\end{equation}
and considered within various cosmological models to reveal some cosmological consequences~\cite{Martiros2}-\cite{Martiros5} of it. $\alpha$ and $\beta$ are dimensionless constants, the energy density $\rho$ could be $\rho_{m}$, $\rho_{\small{de}}$, $\rho_{tot}$. $q$ is the deceleration parameter
\begin{equation}\label{eq:decparameter}
q=-\frac{1}{H^{2}} \frac{\ddot{a}}{a}=-1-\frac{\dot{H}}{H^{2}}.
\end{equation}
This new type of interaction can change its sign when our Universe changes from the deceleration $q>0$ to the acceleration $q<0$ Universe. It should be noted that the sign changeability for the interaction other approaches also could be used. Recently, other tendency concerning to the phenomenological modification of the interaction terms was observed, for instance, such that the constants $b$ and $\gamma$ from interaction terms $Q$ assumed to be function of the scale factor $b(a)=b_{0}a^{\zeta}$~\cite{Saridakis}, which is also an interesting modification with its wide interesting consequences. In this work as a first and a simple model we will consider an Universe with a single component fluid. Then as a generalization of the model we will consider a composed fluid model with different couplings between the fluid components. Examples for the interaction term $Q$ considered in this work could be presented as a particular forms of more general form given as
\begin{equation}\label{eq:generalQ}
Q = q^{n}(3  b a^{\chi} H \rho + \gamma a^{\epsilon}\dot{\rho}),
\end{equation}
where $q$ is the deceleration parameter, $H$ is the Hubble parameter, $a$ is the scale factor and $\rho$ is the energy density of the Universe, while $n$, $b$ and $\gamma$ are constants and should be determined from the observations. \\\\
The paper is organized as followed. After introduction, in the next section we consider a model of the Universe with varying Ghost DE. We will consider composed fluid models in the next section. One of the sections devoted to the observational constraints on the models with the causality issue to eliminate phenomenology from the models where we include analysis of the Generalized Second Law of Thermodynamics. Discussion of obtained results presented in last section.

\section{\large{The Universe with varying Ghost DE}}

\begin{figure}[h]
 \begin{center}$
 \begin{array}{cccc}
 \includegraphics[width=60 mm]{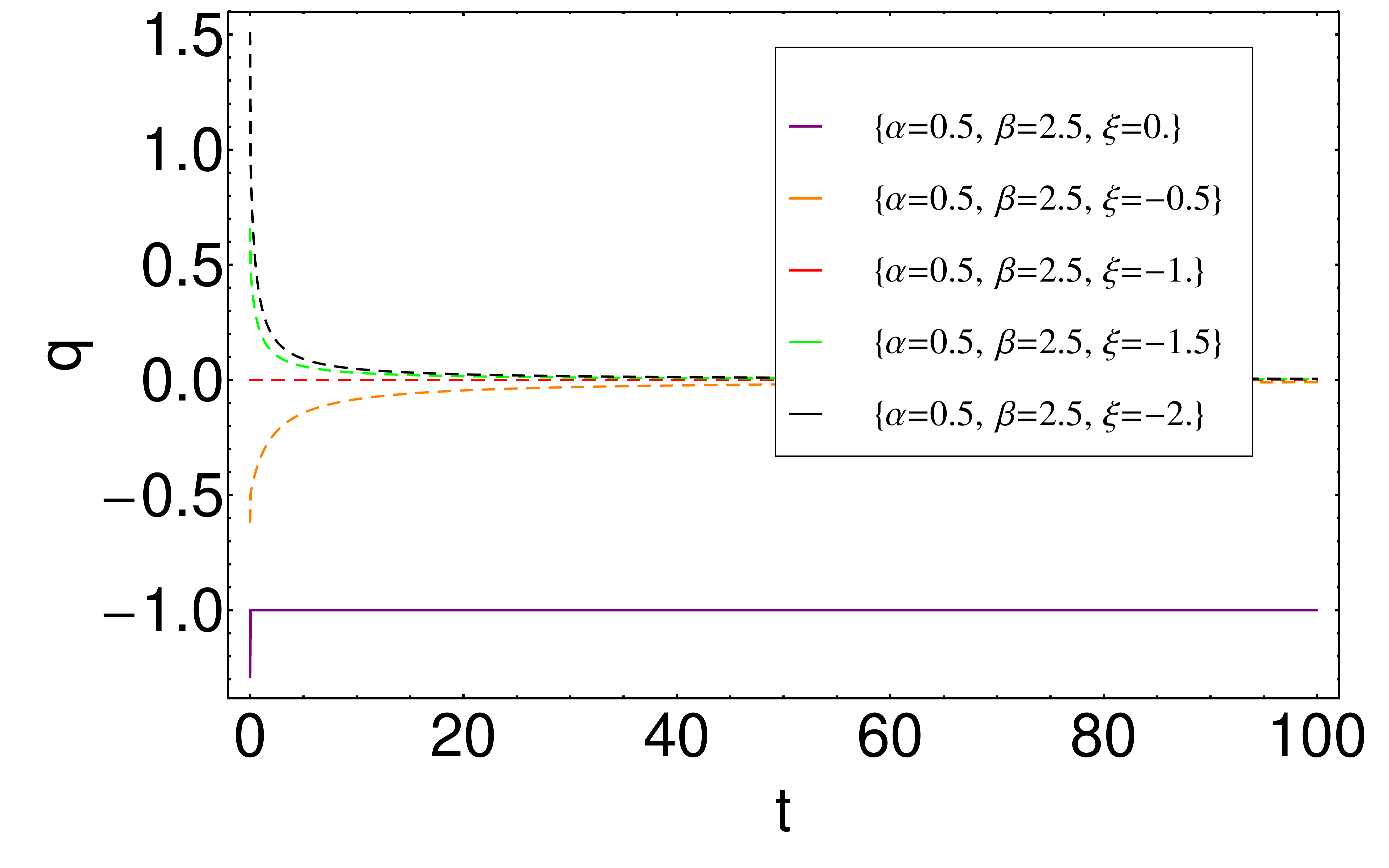} &
\includegraphics[width=60 mm]{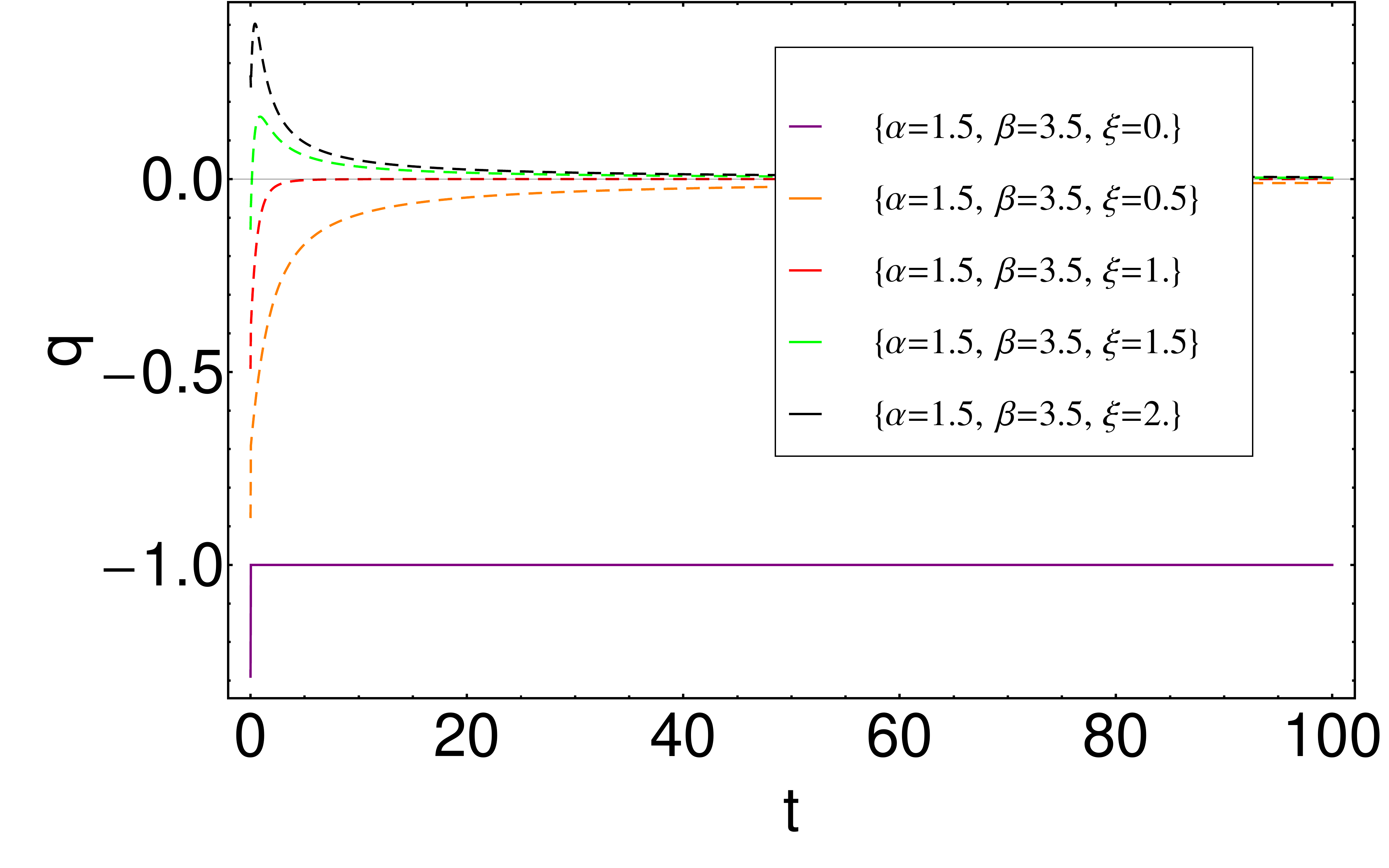} \\
 \end{array}$
 \end{center}
 \caption{Behavior of the deceleration parameter $q$ against time $t$ for the Universe containing only varying Ghost DE for negative and positive $\xi$. $\xi$ is the parameter of the proposed modification. The model with $\xi = 0$ corresponds to the usual Ghost DE with $\rho = \alpha H(t)+ \beta H(t)^{2}$.}
 \label{fig:single_q}
\end{figure}

\begin{figure}[h]
 \begin{center}$
 \begin{array}{cccc}
 \includegraphics[width=63 mm]{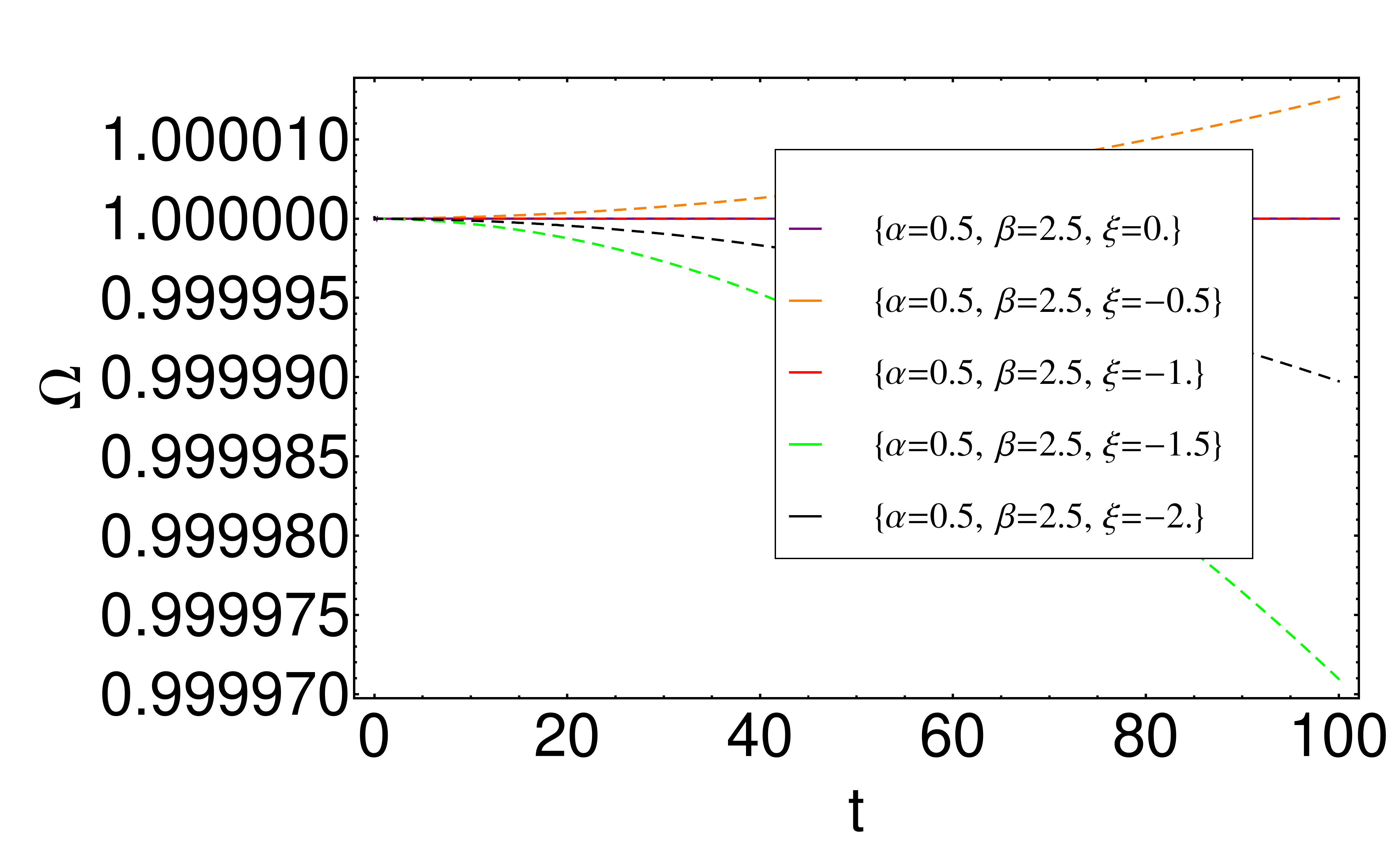} &
\includegraphics[width=60 mm]{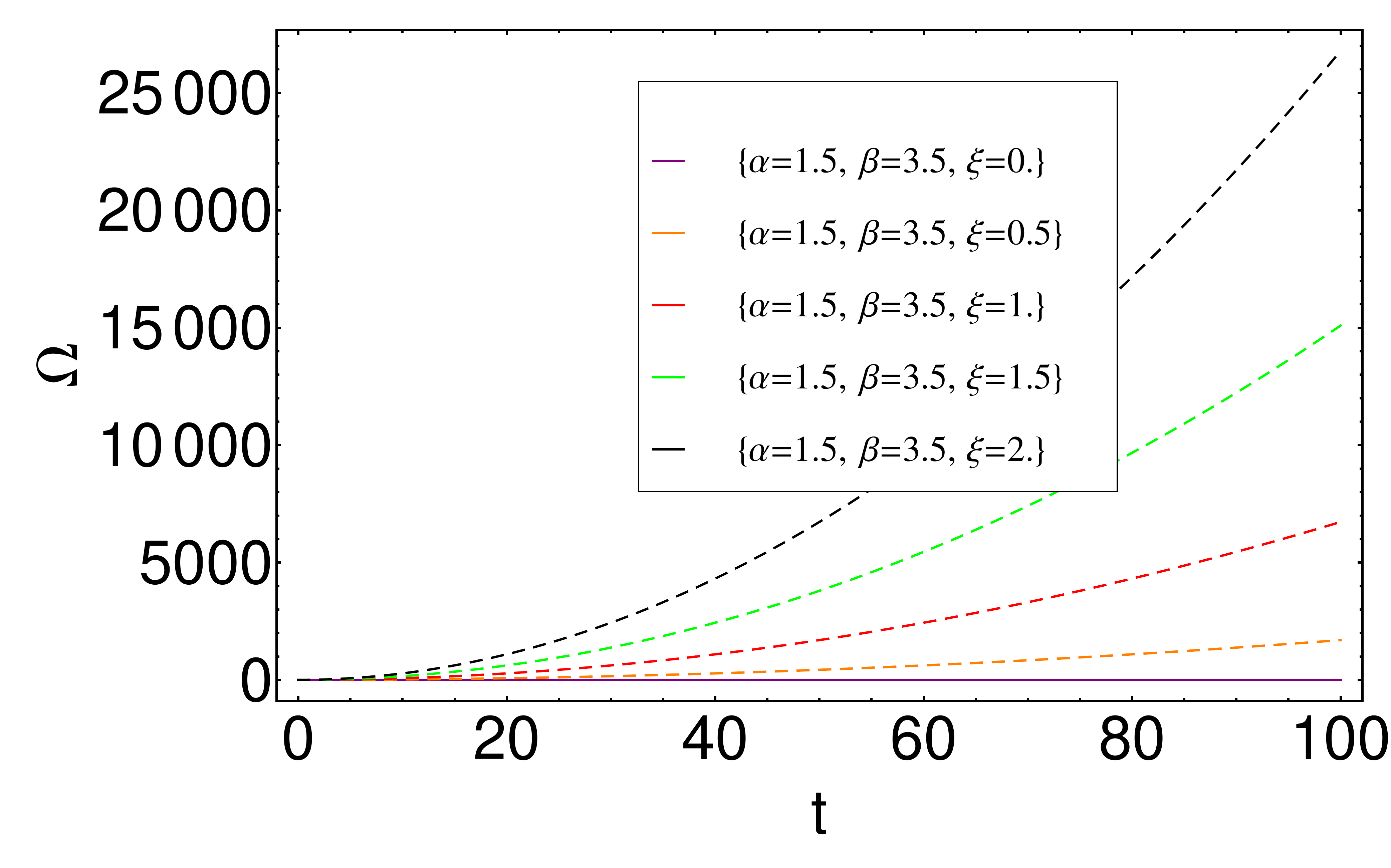} \\
 \end{array}$
 \end{center}
 \caption{$\Omega = \frac{\rho}{3H^{2}}$ of the Universe containing only varying Ghost DE for negative and positive $\xi$. $\xi$ is the parameter of the proposed modification. The model with $\xi = 0$ corresponds to the usual Ghost DE with $\rho = \alpha H(t)+ \beta H(t)^{2}$.}
 \label{fig:single_Omega}
\end{figure}

Modeling the dark sector of the Universe within a varying Ghost DE could be a good starting point to understand some aspects of  the phenomenological modification, therefore we would like to start our analysis of the dynamics of the Universe with a single fluid content. Phenomenological dark energy model of our interest which results from the modification of the energy density of the Ghost DE refers as a varying Ghost DE. Our first attempt is an assumption that the energy density of the DE is given as
\begin{equation}\label{eq:singlemodel}
\rho =\alpha a(t)^{\xi} H(t)+ \beta H(t)^{2},
\end{equation}
where ${\xi}$, $\alpha$ and $\beta$ are constants and $\xi = 0$ will reduce our new model to the original Ghost DE.
The pressure of the De model can be found from the energy conservation equation
\begin{equation}
\dot{\rho} + 3H(\rho+P) = 0,
\end{equation}
and reads as
\begin{equation}\label{eq:psinglemodel}
P = -\frac{\dot{H}}{3} \left [ \frac{\alpha a^{\xi}}{H} +2\beta  \right ] -\frac{(\xi+3)\alpha a^{\xi}H}{3} -\beta H^{2}.
\end{equation}
Taking into account last two equation as well as Eq.~(\ref{eq:fridman2}) we will have a complete set of equations allowing us to study the behavior of the Universe. If the dynamics of the Universe is controlled via varying Ghost DE suggested in this work, then we can have two different scenarios either we have ever accelerated expansion for the whole history or we have the Universe where $q \geq 0$. For the models with $\xi > 0$ for the later stages of the evolution $q \approx 0$. Similar situation can be seen also for the models with negative $\xi$~(Fig.~(\ref{fig:single_q})).
With the analysis of the behavior of the $\Omega = \frac{\rho}{3H^{2}}$ we conclude that the models with $\xi > 0$ are not possible scenarios, because in such models we have continuously increasing $\Omega$, while due to the observational results we have a well  known fact that $\Omega \approx 1$. While the models with negative $\xi$ could work well, because we can obtain $\Omega \approx 1$~(Fig.~(\ref{fig:single_Omega})). Further analysis reveals that a small positive values for $\xi$ also can provide reasonable results. Next, according to the well accepted fact to model dark sector of the Universe with a DE and DM, where usually DM interpreted as a cold dark matter (CDM) with $\omega_{m} = 0$. It is also well known that introduction of an interaction between DE and DM can solve cosmological coincidence problem. Therefore in the next section we will discuss the models involving CDM and an interaction and possible solutions for the cosmological coincidence problem within a particular model of the varying Ghost DE.

\section{\large{Two component effective fluid models}}
\begin{figure}[h]
 \begin{center}$
 \begin{array}{cccc}
 \includegraphics[width=60 mm]{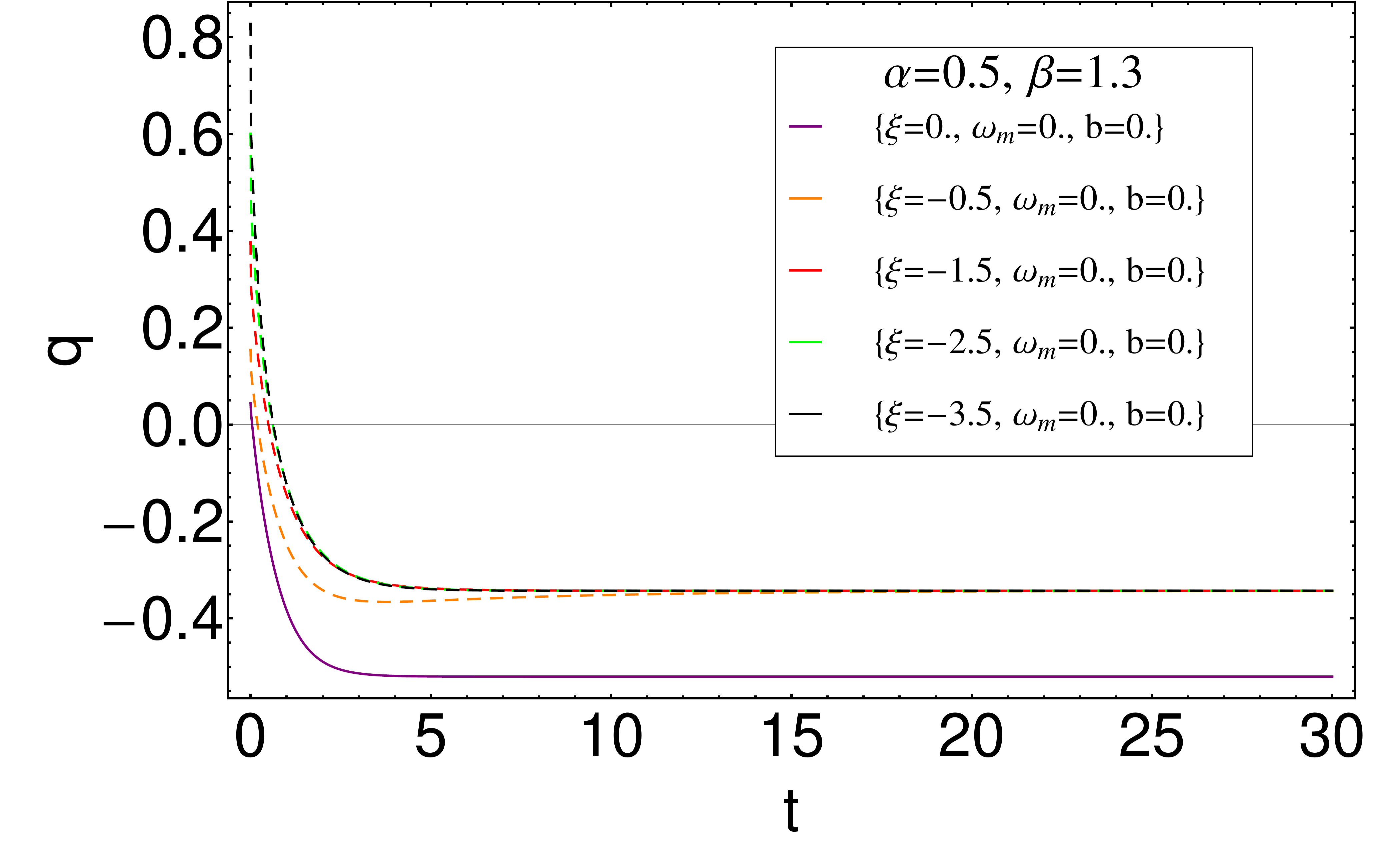} &
\includegraphics[width=60 mm]{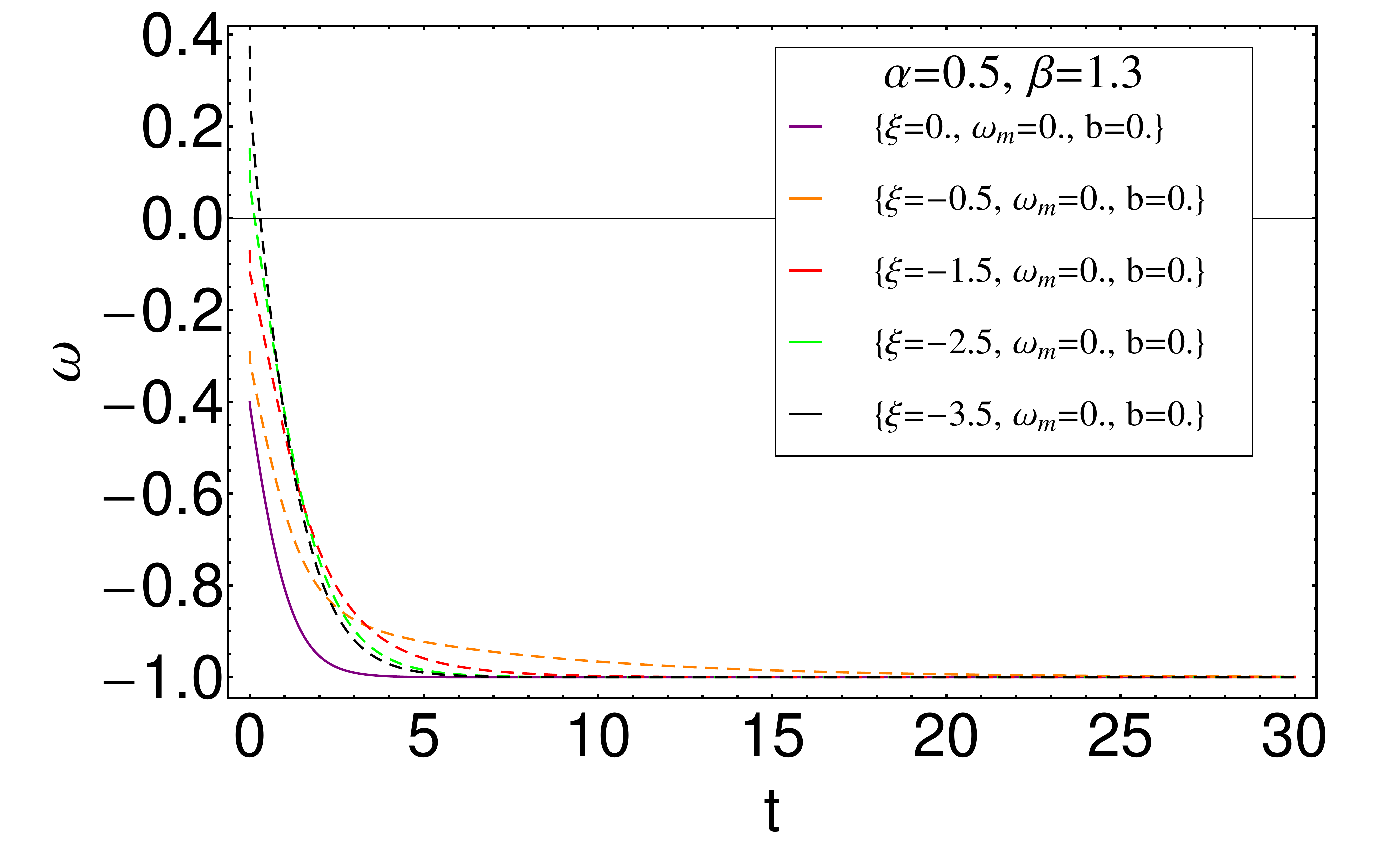} \\
 \end{array}$
 \end{center}
 \caption{Behavior of the deceleration parameter $q$ and EoS parameter $\omega$ of the varying Ghost DE against time $t$ for the Universe containing an effective fluid. $\xi$ is the parameter of the proposed modification. The model with $\xi = 0$ corresponds to the usual Ghost DE with $\rho = \alpha H(t)+ \beta H(t)^{2}$. Non-interacting case.}
 \label{fig:effective_nonint}
\end{figure}

\begin{figure}[h]
 \begin{center}$
 \begin{array}{cccc}
 \includegraphics[width=60 mm]{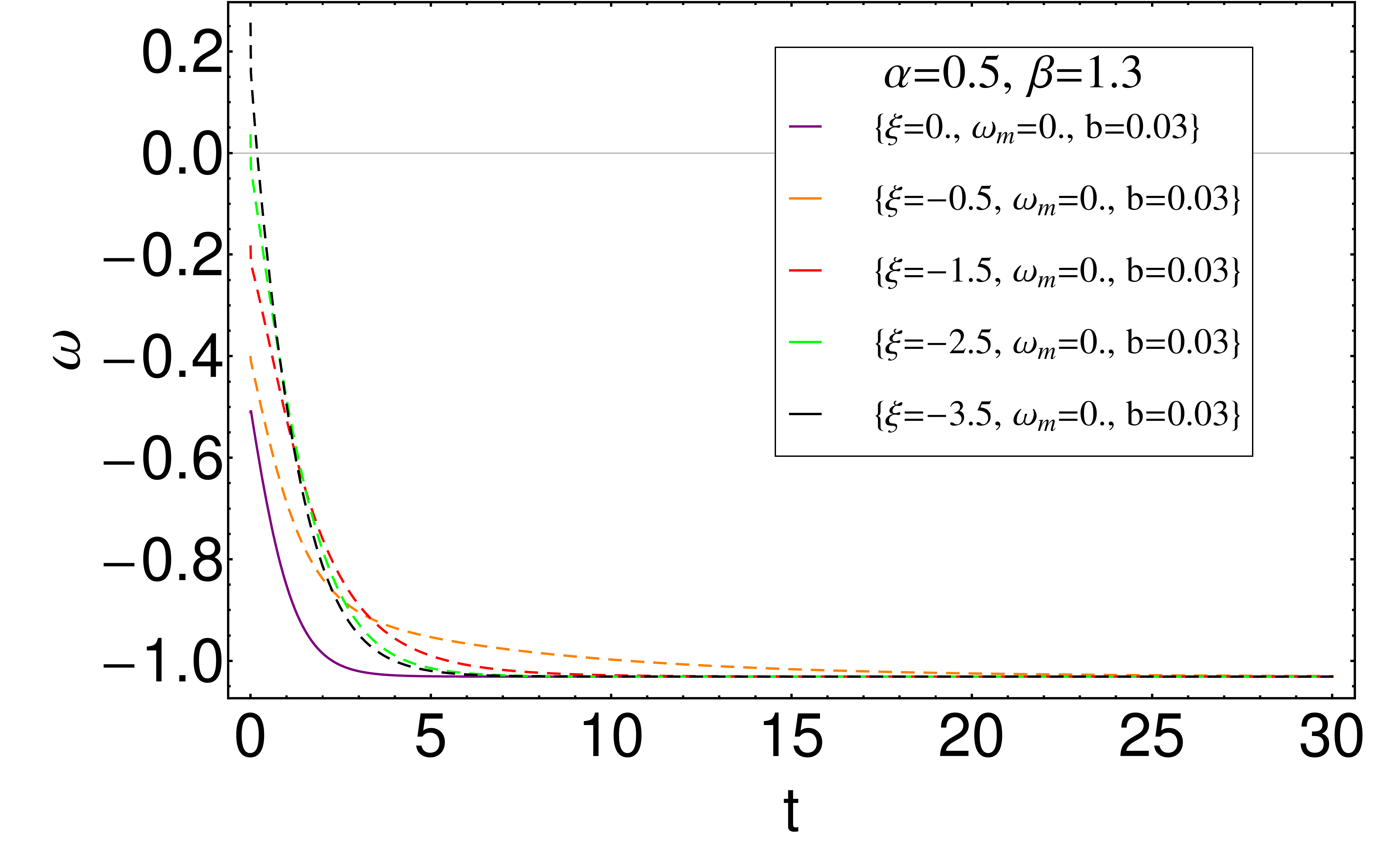} &
\includegraphics[width=60 mm]{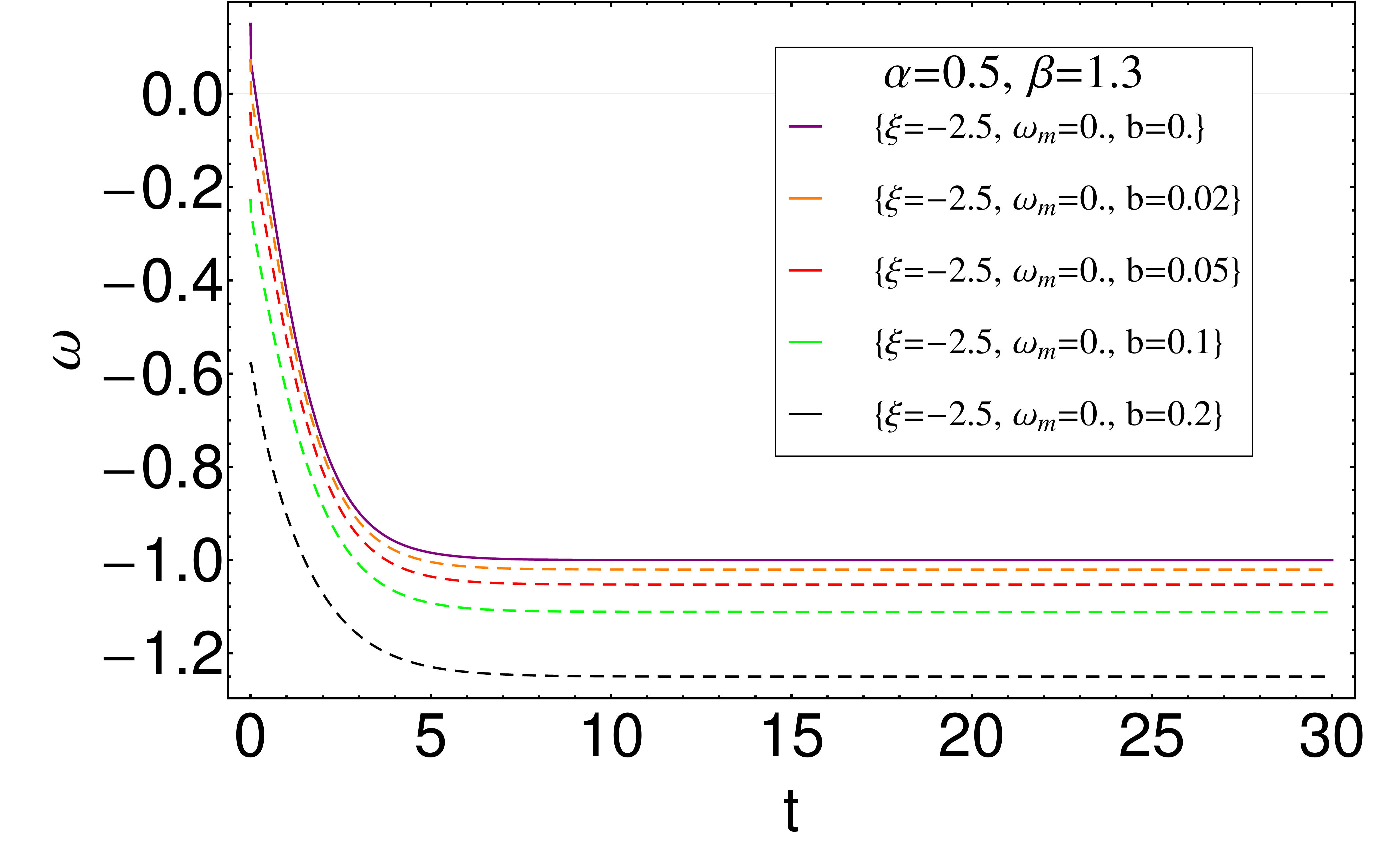} \\
 \end{array}$
 \end{center}
 \caption{Behavior of the EoS parameter $\omega$ of the varying Ghost DE against time $t$ for the Universe containing an effective fluid. $\xi$ is the parameter of the proposed modification. The model with $\xi = 0$ corresponds to the usual Ghost DE with $\rho = \alpha H(t)+ \beta H(t)^{2}$.}
 \label{fig:effective_omega}
\end{figure}

\begin{figure}[h]
 \begin{center}$
 \begin{array}{cccc}
\includegraphics[width=60 mm]{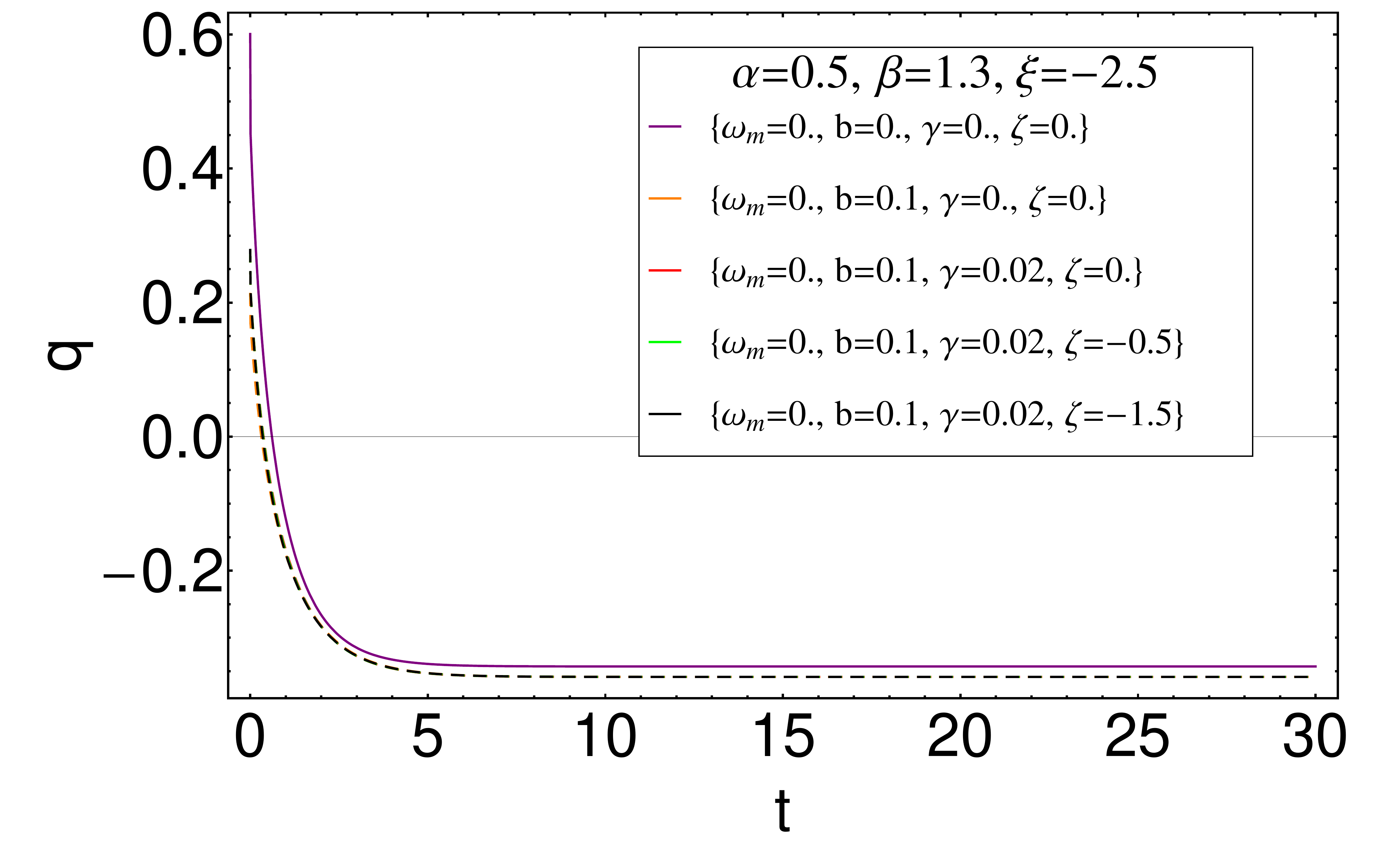} &
\includegraphics[width=60 mm]{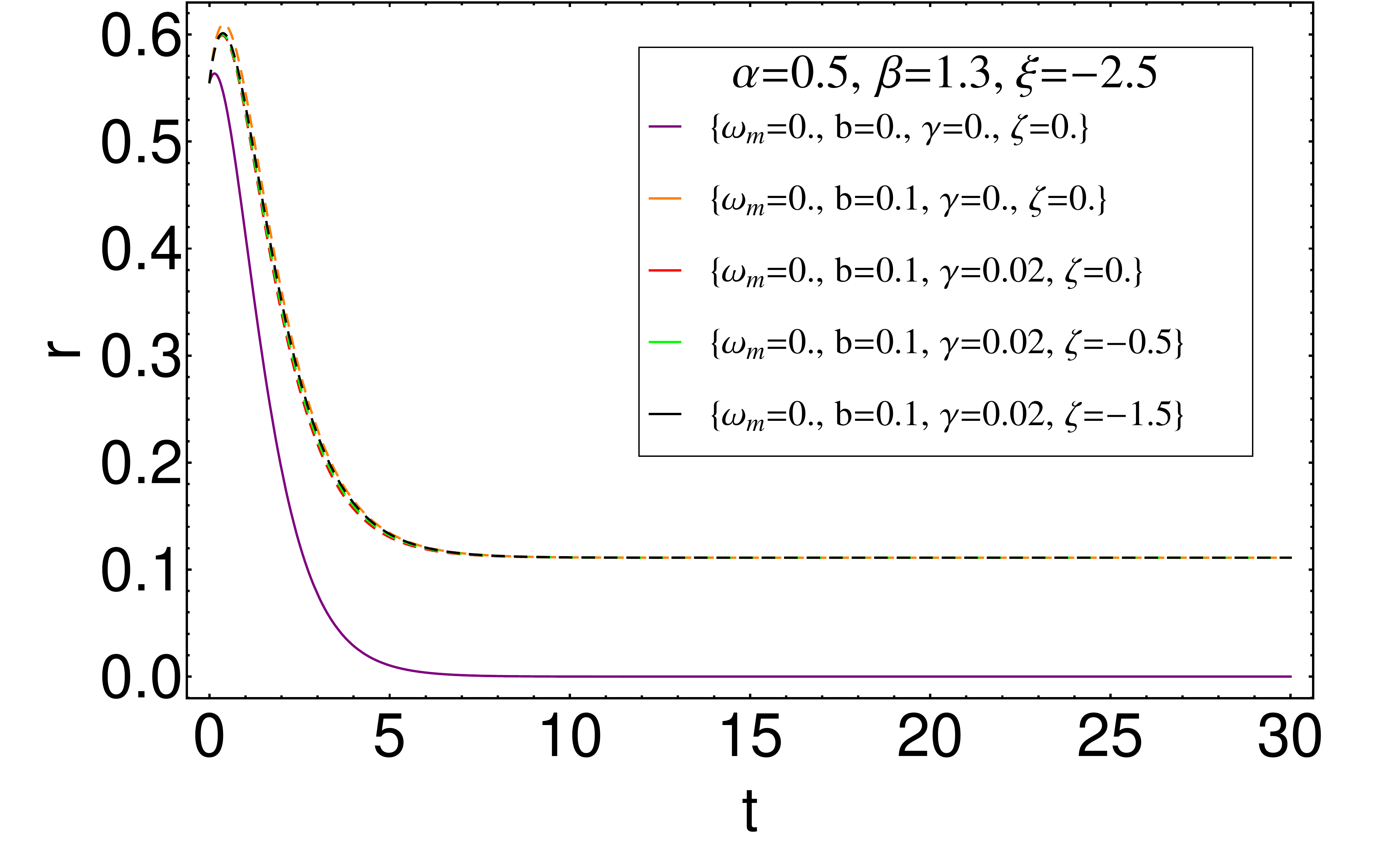} \\
\includegraphics[width=60 mm]{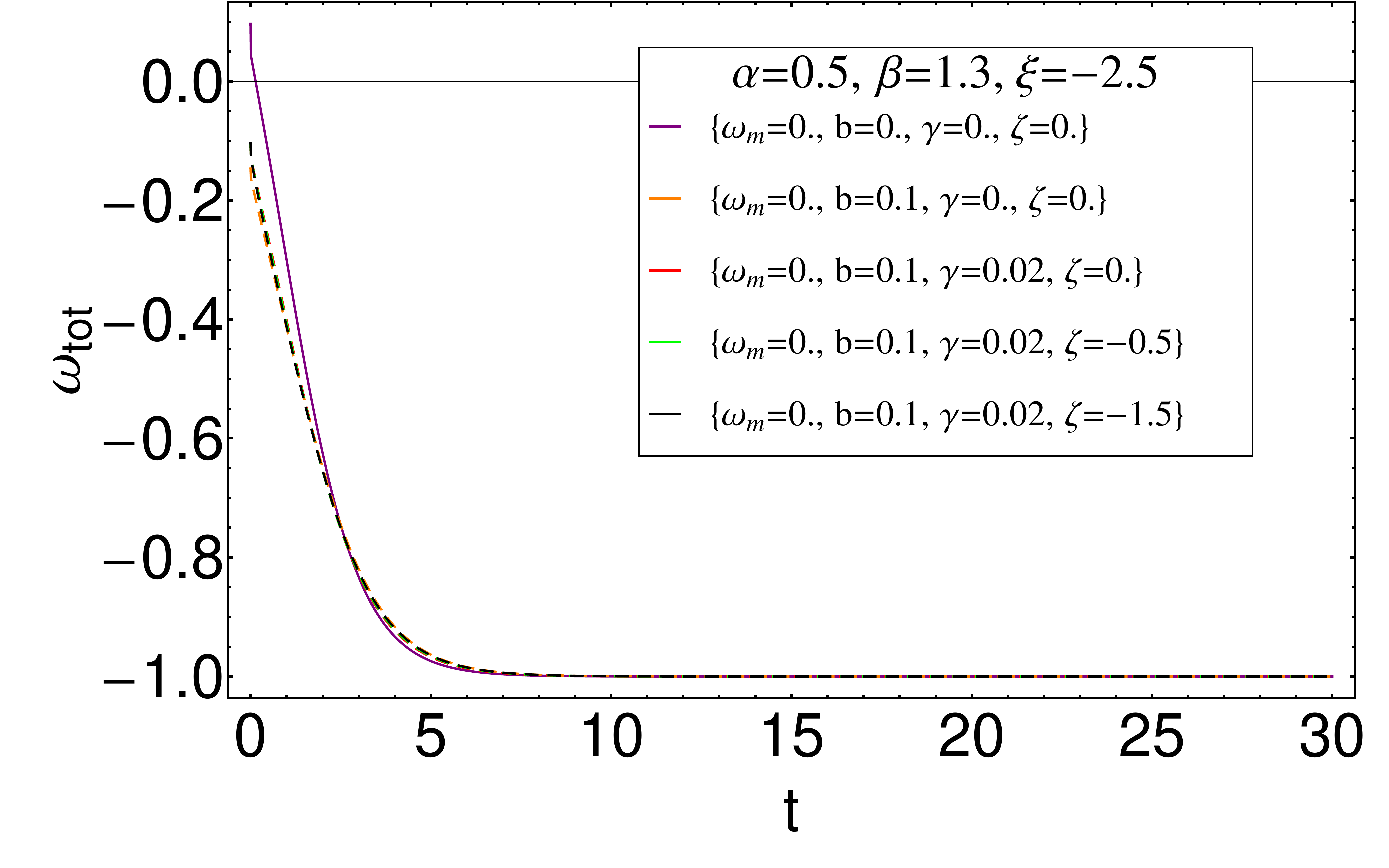} &
\includegraphics[width=60 mm]{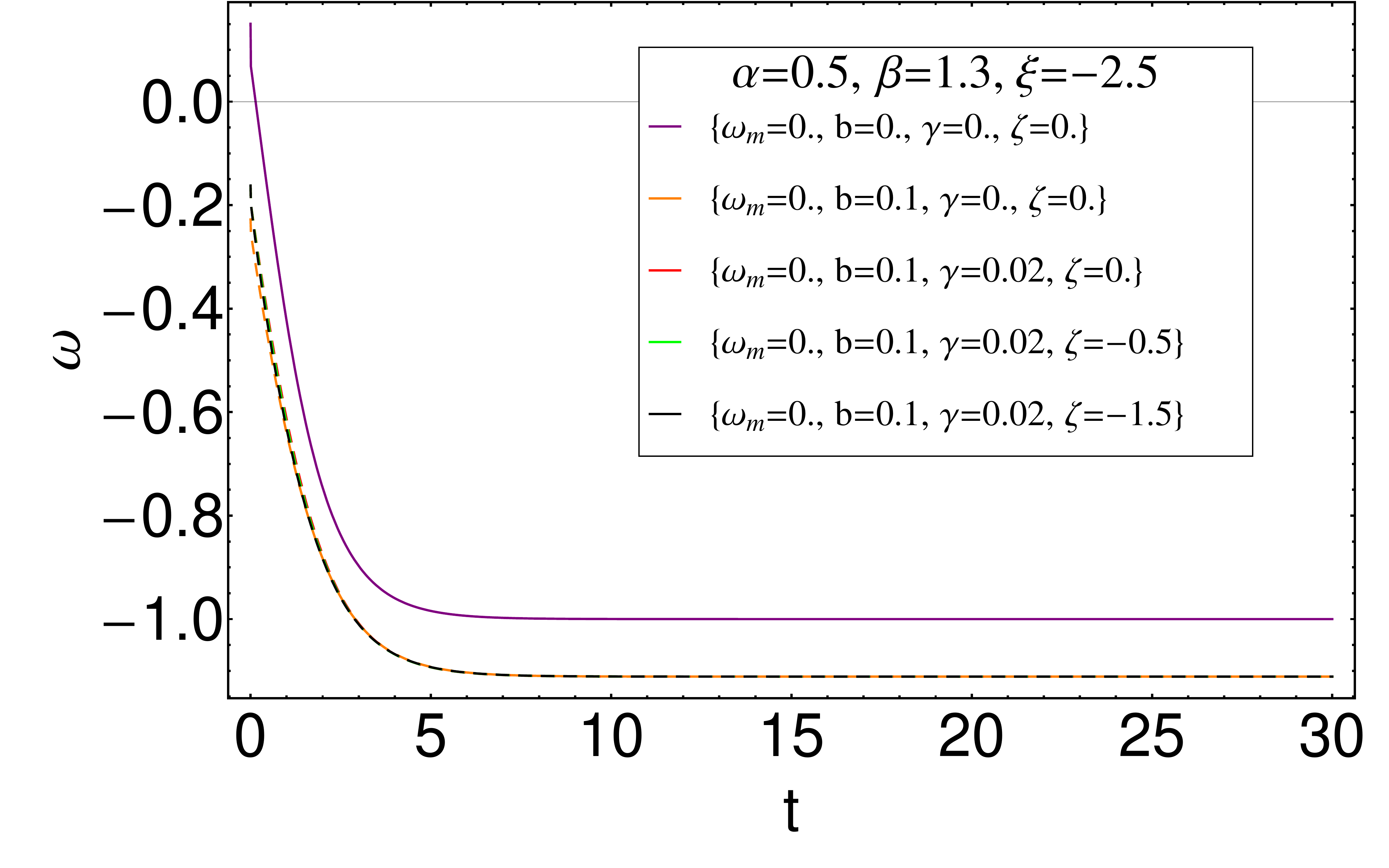} \\
 \end{array}$
 \end{center}
 \caption{Behavior of the deceleration parameter $q$, $r=\frac{\rho_{m}}{\rho_{VG}}$, $\omega_{tot}$ and EoS parameter of the varying Ghost DE $\omega$ against time $t$ for the Universe containing an effective fluid for different forms of the interaction terms $Q$ given by Eq.~(\ref{eq:generalQ}). For all cases the parameter $n=0$. The model with $\xi = 0$ corresponds to the usual Ghost DE with $\rho = \alpha H(t)+ \beta H(t)^{2}$. Blue line corresponds to non interacting case.}
 \label{fig:general}
\end{figure}

The analysis and the results of this section involves models where we have interacting DE and CDM. Cosmography and a possible solution of the cosmological coincidence problem will be under our attention. If we consider non-interacting DE models practically we can handle the problem analytically (for some models), when we have an interaction, then the analysis could be complicated and numerical analysis will be the right tool to understand the behavior of the models in different regimes. Mathematically, in our models when there is not an interaction between the components we will write the energy conservation for the effective fluid with $\rho = \rho_{m} + \rho_{VG}$ and $P=P_{m} + P_{VG}$ as
\begin{equation}\label{eq:nfirstfluid}
\dot{\rho}_{m}+3H(\rho_{m}+P_{m})=0,
\end{equation}
and
\begin{equation}\label{eq:nsecondfluid}
\dot{\rho}_{\small{VG}}+3H(\rho_{\small{VG}}+P_{\small{VG}})=0,
\end{equation}
and for the pressure of the varying Ghost Dark energy to have
\begin{equation}\label{eq:PGDe}
P_{\small{VG}}=\frac{-\dot{\rho}_{\small{VG}}}{3H}-\rho_{\small{VG}}.
\end{equation}
With our modification for the Ghost DE, as we observed, there is a possibility to obtain a transit Universe, where the deceleration parameter changed its sign from the positive to negative providing accelerated expansion observed in the old Universe. This transition becomes apparent with the decreasing the $\xi$ parameter. Another interesting outcome of our phenomenological suggestion is that that with the decreasing of the $\xi$ we are able to change the nature of the Ghost DE in the early Universe. With the decreasing the $\xi$ we will transform DE to a matter like fluid. While independent from the values of the $\xi$ the EoS parameter of the varying Ghost DE will tend to $-1$~(Fig~(\ref{fig:effective_nonint})).
\\\\
Consideration of the interaction $Q$ between the components provide some changes in the mathematics of the problem, particularly the pressure of the varying Ghost DE could be found from
\begin{equation}\label{eq:intPGDe}
P_{\small{VG}}=\frac{-Q-\dot{\rho}_{\small{VG}}}{3H}-\rho_{\small{VG}},
\end{equation}
with appropriate changes in Eq.~({\ref{eq:nfirstfluid}}). One of the forms of the interaction term $Q$ intensively considered in the physical literature is of the form
\begin{equation}\label{eq:classical_int}
Q=3Hb\rho,
\end{equation}
where $H$ is the Hubble parameter, $b$ is the constant and $\rho$ is the energy density of the effective fluid. When we have ever accelerated Universe with the interaction between Ghost DE and CDM given by ~Eq.~(\ref{eq:classical_int}), then with the proposed varying Ghost DE we can see a possibility to have a transit Universe i.e. the transition from the $q > 0$ to $q<0$.
\begin{figure}[h]
 \begin{center}$
 \begin{array}{cccc}
 \includegraphics[width=60 mm]{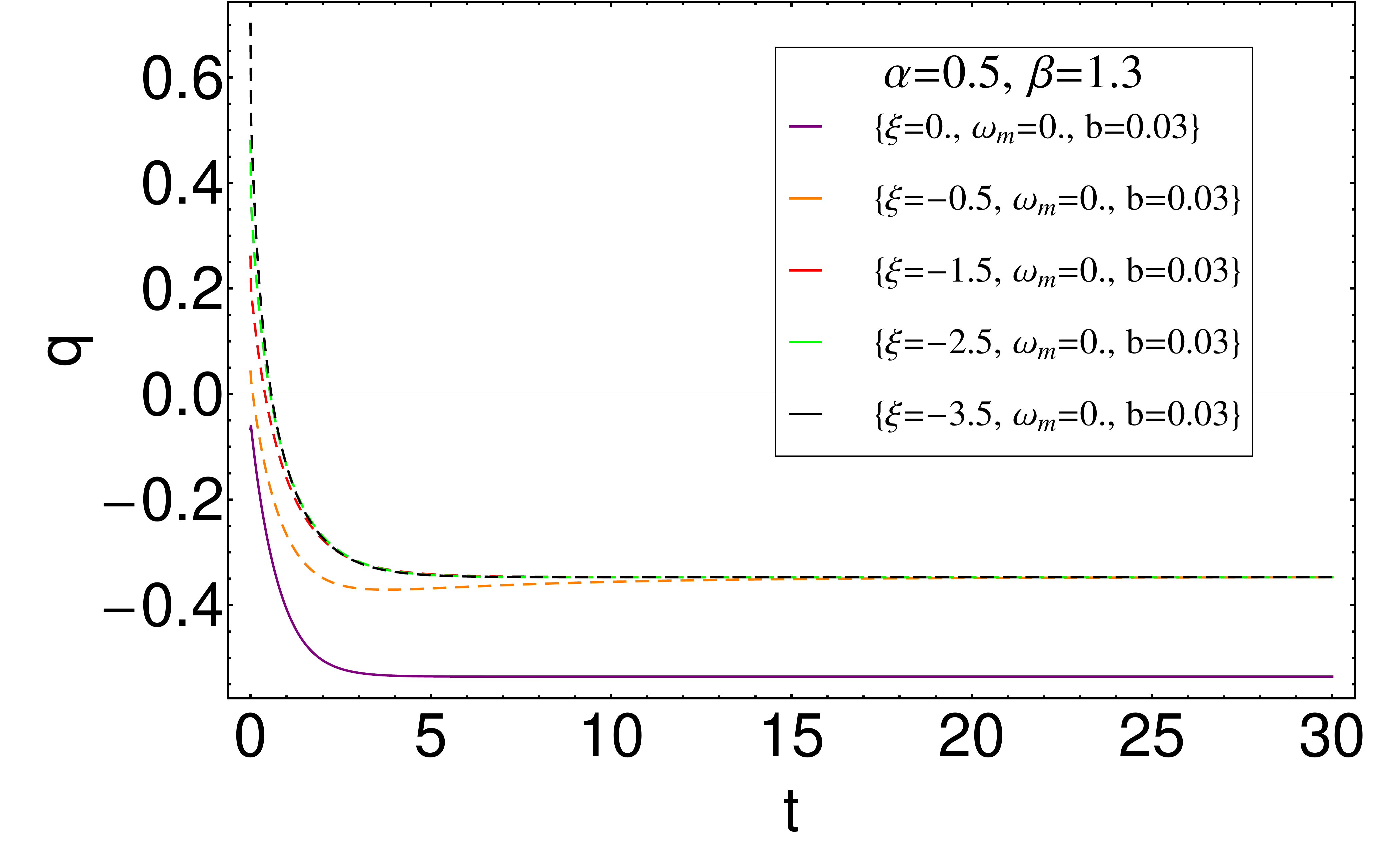} &
\includegraphics[width=60 mm]{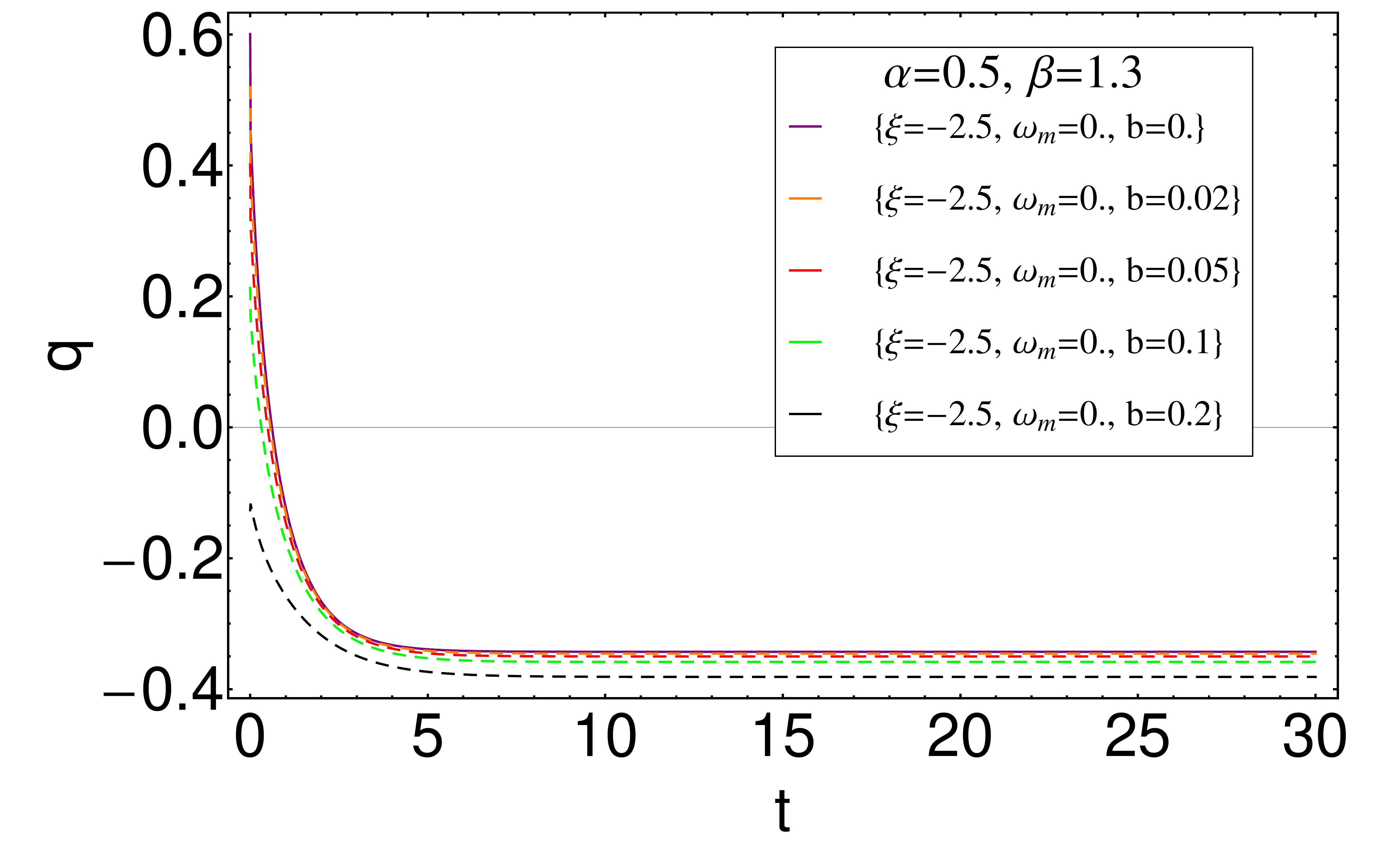} \\
 \end{array}$
 \end{center}
 \caption{Behavior of the deceleration parameter $q$ against time $t$ for the Universe containing an effective fluid. $\xi$ is the parameter of the proposed modification. The model with $\xi = 0$ corresponds to the usual Ghost DE with $\rho = \alpha H(t)+ \beta H(t)^{2}$.}
 \label{fig:effective_q}
\end{figure}
This transition corresponds to the decreasing of $\xi$ parameter. However, we should note that the increasing the value of the interaction parameter $b$ for a fixed value of $\xi$ can provide us an Universe which is also ever accelerated~(Fig.~\ref{fig:effective_q}), therefore, eventually we should apply observational constraints on the models to illuminate correct values of the parameters. Such behavior was obtained for the fixed values of the parameters $\alpha$ and $\beta$. In such Universes EoS parameter of the effective fluid described by
\begin{equation}\label{eq:EOS_eff}
\omega_{tot} = \frac{P_{VG} }{\rho_{m} + \rho_{VG}},
\end{equation}
behaves as a cosmological constant $\omega_{tot}\to 1$ in later stages of the evolution. Decreasing $\xi$ will give usual matter properties to the varying Ghost DE in the early stages of the evolution, but for the latter stages of the evolution independent of the values of the main parameter $\xi$ varying Ghost DE will mimic the cosmological constant with $\omega = -1$. The interaction under our consideration has another interesting effect on the behavior of the EoS parameter $\omega$ of the varying Ghost DE. From the left plot of the Fig.~(\ref{fig:effective_omega}) it becomes clear that with the fixed value of $\xi$ and with increasing the value of $b$ varying Ghost DE can be interpreted as a phantom DE.
Long standing puzzle known as the cosmological coincidence problem can be solved using different approaches. One of the approaches is the consideration of the interaction between dark sector of the Universe. We need another analysis for the deep understanding of the differences between our suggested model and original Ghost DE model, which involves the analysis of the
\begin{equation}
r=\frac{\rho_{m}}{\rho_{VG}},
\end{equation}
which shows us that without interaction this model is not able to solve the cosmological coincidence problem, while with the interaction $Q=3Hb\rho$ we have a solution.
\begin{figure}[h]
 \begin{center}$
 \begin{array}{cccc}
 \includegraphics[width=60 mm]{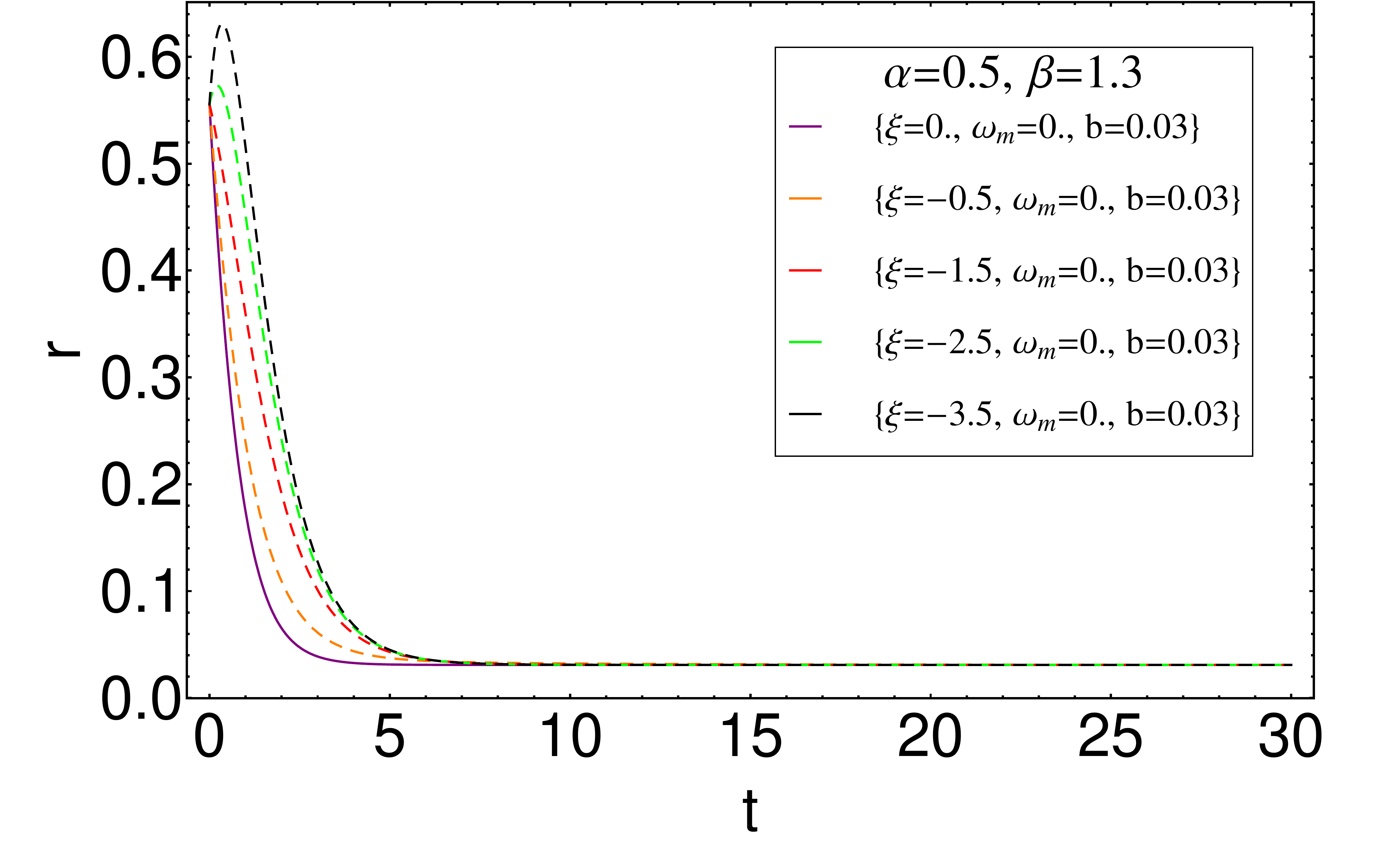} &
\includegraphics[width=60 mm]{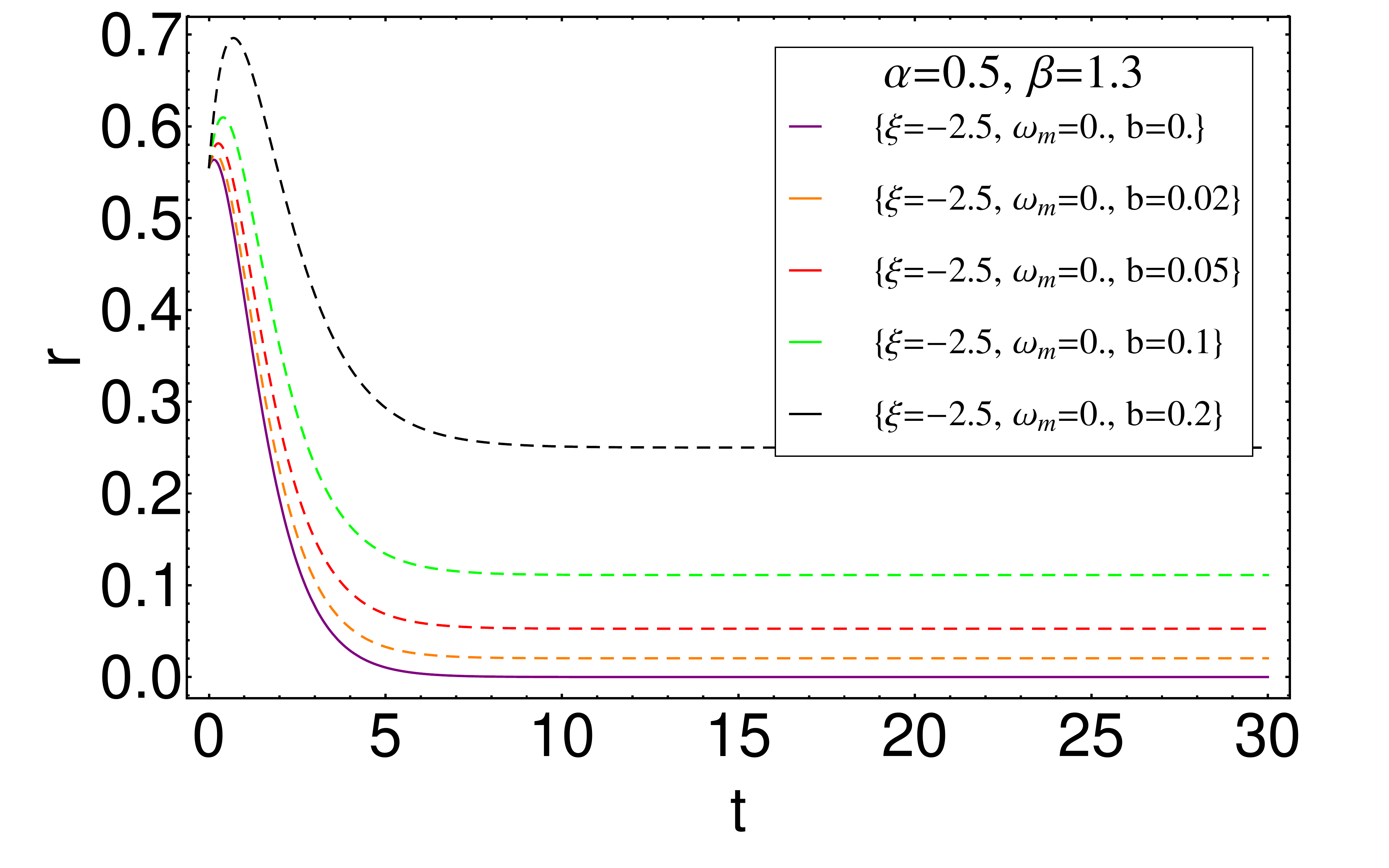} \\
 \end{array}$
 \end{center}
 \caption{Behavior of the $r=\frac{\rho_{m}}{\rho_{VG}}$ against time $t$ for the Universe containing an effective fluid. Plots indicate a possibility to solve the cosmological coincidence problem with $r\to r_{0}$. $\xi$ is the parameter of the proposed modification. The model with $\xi = 0$ corresponds to the usual Ghost DE with $\rho = \alpha H(t)+ \beta H(t)^{2}$.}
 \label{fig:effective_r}
\end{figure}
Moreover we observed that with increasing interaction parameter $b$ evidence of the solution in the form of the scaling solutions is appears with $r\to r_{0}$, where $r_{0}$ is a constant. In the last part of this section we would like to discuss impact of the different forms of the interaction terms $Q$ on our model. We see that with great accuracy obtained results for the different forms for $Q$ given by Eq.(\ref{eq:generalQ}) with $n=0$ are coincide with the results obtained for the interaction term of the $Q = 3Hb\rho$~(Fig.~(\ref{fig:general})) form.

\section{\large{Observational Constraints, Causality issue and The Generalized Second Law of Thermodynamics }}

\begin{figure}[h!]
 \begin{center}$
 \begin{array}{cccc}
\includegraphics[width=70 mm]{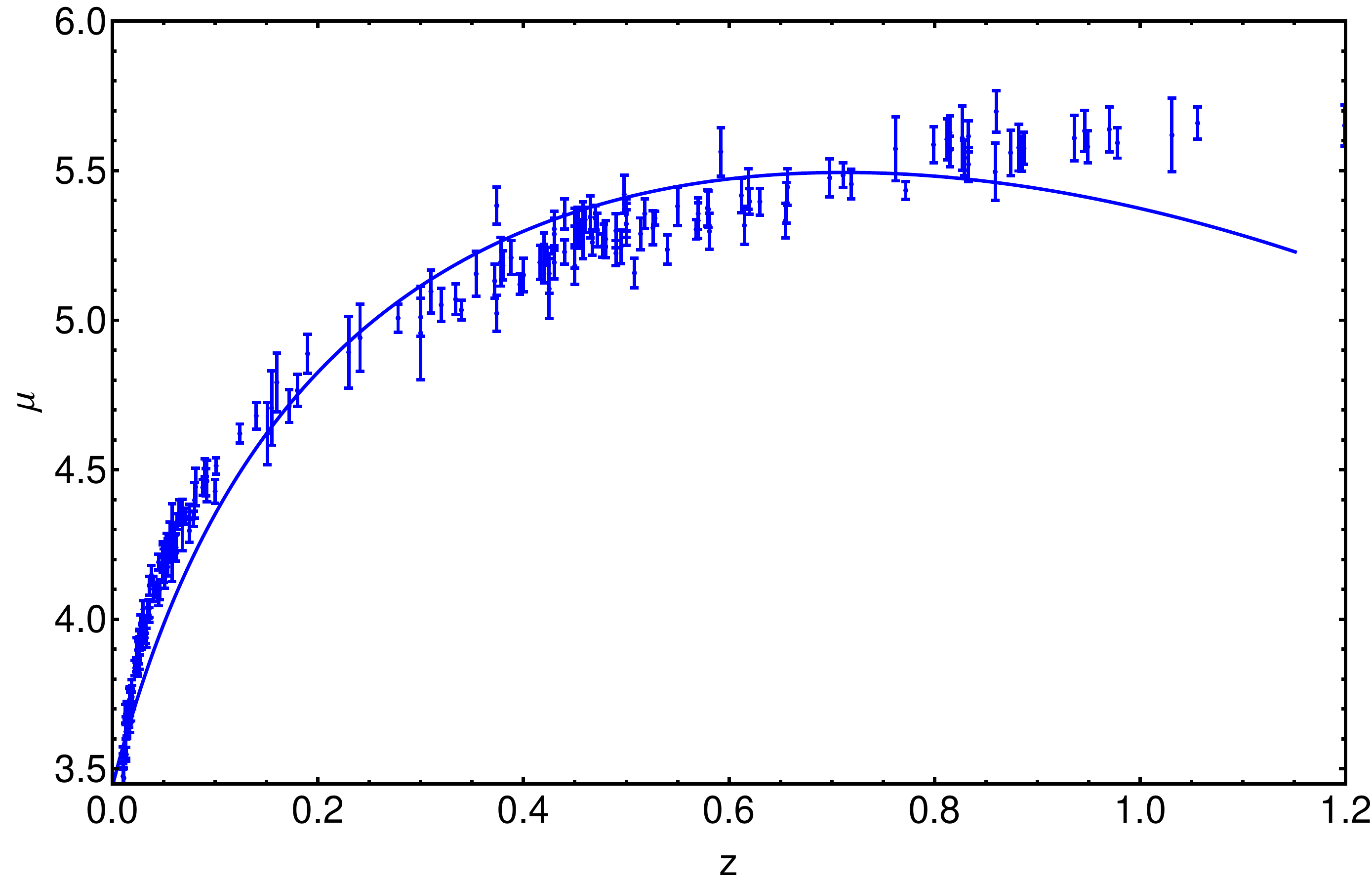}
 \end{array}$
 \end{center}
\caption{Observational data SneIa + BAO + CMB for distance modulus versus our theoretical results.}
 \label{fig:modelobs}
\end{figure}

Comparison of a theoretical model with observational data is a good way to understand the validity of the theoretical model as well as it is a powerful tool to illuminate physical aspects of the model from the phenomenological one. Besides finding best fit of the model with the observations another simply way exist which allows to reject or accept the models and it is the square of the sound speed defined as
\begin{equation}\label{eq:soundspeed}
C_{s}^{2} = \frac{\partial{P}}{\partial{\rho}},
\end{equation}
where in our case $P$ is the pressure of the effective fluid and $\rho$ is the energy density of the same effective fluid. We have well defined range for the square of the sound speed which is
\begin{equation}\label{eq:CS2range}
0 \leq C_{s}^{2} \leq 1,
\end{equation}
either to accept or reject the theoretical models. In our models we saw that with $C_{s}^{2} \to 0$ in the latter stages of the evolution, therefore our models could be appropriate models of the old Universe if we follow to the widespread accepted opinion. But if we follow to the ideas challenging Eq.~(\ref{eq:CS2range}) then we have good chances to extend proposed models of this work and obtain other interesting behaviors differ than presented in this work. Comparing theoretical models with observational data we found a good fit of the models up to $z \approx 0.9$~(Fig.~(\ref{fig:modelobs}))

\begin{table}
  \centering
    \begin{tabular}{ | l | l | l | l | l | l | l | p{5cm} |}
    \hline
    $Model$ & $\alpha$ & $\beta$ & $b$ & $\xi$  \\
  \hline
   $Q = 3Hb\rho$ & $0.5^{+0.25}_{-0.15}$& $1.3^{+0.2}_{-0.1}$ & $ 0.02^{+0.02}_{-0.01}$ & $ 0.01^{+0.03}_{-0.28}$ \\
    \hline

    \end{tabular}
\caption{ Values of the model parameters obtained from the $SneIa + BAO + CMB$ data  for distance modulus versus theoretical results for the two-component fluid universe with varying Ghost DE. }
  \label{tab:2Table}
\end{table}
Another important question is also the validity of the Generalized Second Law of Thermodynamics. The foundation of GSL required the Gibb’s equation of thermodynamics is
\begin{equation}\label{Gibs}
  T_{X}dS_{IX}= PdV_{X} + dE_{IX}
\end{equation}
where $S_{IX}$ and $E_{IX}=\rho V_{X}$, are  internal entropy and energy within the horizon,  while $V_{X}=\frac{4}{3}\pi R^{3}_{X}$ be the volume of sphere with horizon radius \[R_{X}=\left(\sqrt{H^{2}+\frac{k}{a^{2}}}\right)^{-1}. \]. In order the GSL to be hold it is required that $\dot{S}_{X}+\dot{S}_{IX}\geq0$ i.e. the sum of entropy of matter enclosed by horizon must be not be a decreasing function of time. Following the work~\cite{Ujjal}where was considered validity of the Generalized Second Law of Thermodynamics for the Universe bounded by the Hubble horizon
\begin{equation}\label{eq:Habblehor}
R_{H}=\frac{1}{H},
\end{equation}
cosmological event horizon
\begin{equation}\label{eq:cosevhor}
R_{E}= a\int_{t}^{\infty}\frac{dt}{a},
\end{equation}
and the particle horizon
\begin{equation}\label{eq:particlehor}
R_{P}=a\int_{0}^{t}\frac{dt}{a},
\end{equation}
we found that the validity of the Generalized Second Law of Thermodynamics for our Universe bounded by the Hubble horizon is also satisfied. Recall that GSL with First Law for the time derivative of total entropy gives
\begin{equation}\label{eq:UF}
\dot{S}_{X}+\dot{S}_{IX}=\frac{R^{2}_{X}}{GT_{X}}\left(\frac{k}{a^{2}}-\dot{H}\right)\dot{R}_{X}.
\end{equation}
while in case without First Law used we get
\begin{equation}\label{eq:UNUF}
\dot{S}_{X}+\dot{S}_{IX}= \frac{2\pi R_{X}}{G}\left[ R^{2}_{X}\left(\frac{k}{a^{2}}-\dot{H}\right)(\dot{R}_{X}-HR_{X})+\dot{R}_{X} \right].
\end{equation}
Under the notations used above we understood that $T_{X}=\frac{1}{2\pi R_{X} }$ and $R_{X}$ is temperature and Radius for a given horizon under equilibrium thermodynamics respectively, $S_{X}$ is the horizon entropy and $\dot{S}_{IX}$ as the rate of change of internal entropy.

\section*{\large{Discussion}}
In this article we proposed and considered a varying Ghost Dark energy. In base of the generalized Ghost dark energy with energy density $\rho_{\small{GDe}}=\alpha H + \beta H^{2}$ we assume that $\alpha$ can be a function of the scale factor, for instance, $\alpha(a)=a^{\xi}$ of the form. The origin of such fluid assumed to be phenomenological. We analyzed the behavior of the Universe modeling the dark sector of it within proposed fluid and observed that having negative values of the parameter $\xi$ is favorable due to the fact that only in that case we have $\Omega \approx 1$. The modeling of the dark sector of the Universe via interacting varying Ghost DE and cold DM is considered as a realistic scenario, therefore we investigated the dynamics of the Universe from this part also. In this case we observed a transit Universe, we observed also that varying Ghost Dark energy behaves as a matter like fluid in the early Universe, while it can be either a cosmological constant or phantom like DE depends on the interplay between parameters $\xi$and $b$. The effective fluid is always a cosmological constant with $\omega \to -1$ with $t \to \infty$. We also studied the behavior of the Universe in case of different interactions between varying Ghost DE and CDM and conclude that with great accuracy obtained results interpret the results obtained of the interaction of the form $Q=3Hb\rho$. From the observational data we found the values of the parameters giving us the best fit of our theoretical results with observations. We also found that the causality issue is satisfied for our models. As the last step we check the validity of the Generalized Second Law of the Thermodynamics for the Universe bounded by the Hubble horizon and found it to be satisfied. We also saw that the proposed modification is also a way to solve the cosmological coincidence puzzle.

\newpage


\begin{thebibliography}{1}

\bibitem{Riess}
A.G. Riess et al., Astron. J. 116 1009 (1998)

\bibitem{Riess1}
S Perlmutter et al., Astrophys. J. 517, 565 (1999)

\bibitem{Riess2}
R. Amanullah et al., Astrophys. J. 716, 712 (2010)

\bibitem{Pope}
A.C. Pope et al. Astrophys. J. 607 655 (2004)

\bibitem{Spergel}
D.N. Spergel et al. Astrophys. J. Supp. 148 175 (2003)

\bibitem{Steinhardt}
P.J. Steinhardt, Critical Problems in Physics, Prinston University Press  (1997)

\bibitem{Sola}
J. Sola and H. Stefancic, Phys. Lett. B 624, 147 (2005)

\bibitem{Ratra}
B. Ratra and P. J. E. Peebles, Phys. Rev. D 37, 3406 (1988)

\bibitem{Ratra1}
C. Wetterich, Nucl. Phys. B 302, 668 (1988)

\bibitem{Ratra2}
A. R. Liddle and R. J. Scherrer, Phys. Rev. D 59, 023509 (1999)
\bibitem{Ratra3}
I. Zlatev, L. M. Wang and P. J. Steinhardt, Phys. Rev. Lett. 82, 896 (1999)
\bibitem{Ratra4}
Z. K. Guo, N. Ohta and Y. Z. Zhang, Mod. Phys. Lett. A 22, 883 (2007)
\bibitem{Ratra5}
S. Dutta, E. N. Saridakis and R. J. Scherrer, Phys. Rev. D 79, 103005 (2009)
\bibitem{Ratra6}
E. N. Saridakis and S. V. Sushkov, Phys. Rev. D 81, 083510 (2010)

\bibitem{Caldwell}
R. R. Caldwell, M. Kamionkowski and N. N. Weinberg, Phys. Rev. Lett. 91, 071301 (2003)
\bibitem{Caldwell1}
R. R. Caldwell, Phys. Lett. B 545, 23 (2002)
\bibitem{Caldwell2}
S. Nojiri and S. D. Odintsov, Phys. Lett. B 562, 147 (2003)
\bibitem{Caldwell3}
 P. Singh, M. Sami and N. Dadhich, Phys. Rev. D 68, 023522 (2003)
\bibitem{Caldwell4}
 J. M. Cline, S. Jeon and G. D. Moore, Phys. Rev. D 70, 043543 (2004)
\bibitem{Caldwell5}
V. K. Onemli and R. P. Woodard, Phys. Rev. D 70, 107301 (2004)
\bibitem{Caldwell6}
W. Hu, Phys. Rev. D 71, 047301 (2005)
\bibitem{Caldwell7}
M. R. Setare and E. N. Saridakis, JCAP 0903, 002 (2009)
\bibitem{Caldwell8}
 E. N. Saridakis, Nucl. Phys. B 819, 116 (2009)
\bibitem{Caldwell9}
 S. Dutta and R. J. Scherrer, Phys. Lett. B 676, 12 (2009)

\bibitem{Feng}
B. Feng, X. L. Wang and X. M. Zhang, Phys. Lett. B 607, 35 (2005)
\bibitem{Feng1}
E. Elizalde, S. Nojiri and S. D. Odintsov, Phys. Rev. D 70, 043539 (2004)
\bibitem{Feng2}
Z. K. Guo, et al., Phys. Lett. B 608, 177 (2005)
\bibitem{Feng3}
M.-Z Li, B. Feng, X.-M Zhang, JCAP, 0512, 002 (2005)
\bibitem{Feng4}
B. Feng, M. Li, Y.-S. Piao and X. Zhang, Phys. Lett. B 634, 101 (2006)
\bibitem{Feng5}
S. Capozziello, S. Nojiri and S. D. Odintsov, Phys. Lett. B 632, 597 (2006)
\bibitem{Feng6}
W. Zhao and Y. Zhang, Phys. Rev. D 73, 123509 (2006)
\bibitem{Feng7}
Y. F. Cai, T. Qiu, Y. S. Piao, M. Li and X. Zhang, JHEP 0710, 071 (2007)
\bibitem{Feng8}
E. N. Saridakis and J. M. Weller, Phys. Rev. D 81, 123523 (2010)
\bibitem{Feng9}
Y. F. Cai, T. Qiu, R. Brandenberger, Y. S. Piao and X. Zhang, JCAP 0803, 013 (2008)
\bibitem{Feng10}
M. R. Setare and E. N. Saridakis, Phys. Lett. B 668, 177 (2008)
\bibitem{Feng11}
M. R. Setare and E. N. Saridakis, Int. J. Mod. Phys. D 18, 549 (2009)
\bibitem{Feng12}
Y. F. Cai, E. N. Saridakis, M. R. Setare and J. Q. Xia, Phys. Rept. 493, 1 (2010)
\bibitem{Feng13}
T. Qiu, Mod. Phys. Lett. A 25, 909 (2010).

\bibitem{Hsu}
S. D. H. Hsu, Phys. Lett. B 594, 13 (2004)
\bibitem{Hsu1}
M. Li, Phys. Lett. B 603, 1 (2004)
\bibitem{Hsu2}
Q. G. Huang and M. Li, JCAP 0408, 013 (2004)
\bibitem{Hsu3}
M. Ito, Europhys. Lett. 71, 712 (2005)
\bibitem{Hsu4}
X. Zhang and F. Q. Wu, Phys. Rev. D 72, 043524 (2005)
\bibitem{Hsu5}
D. Pavon and W. Zimdahl, Phys. Lett. B 628, 206 (2005)
\bibitem{Hsu6}
S. Nojiri and S. D. Odintsov, Gen. Rel. Grav. 38, 1285 (2006)
\bibitem{Hsu7}
E. Elizalde, S. Nojiri, S. D. Odintsov and P. Wang, Phys. Rev. D 71, 103504 (2005)
\bibitem{Hsu8}
H. Li, Z. K. Guo and Y. Z. Zhang, Int. J. Mod. Phys. D 15, 869 (2006)
\bibitem{Hsu9}
E. N. Saridakis, Phys. Lett. B 660, 138 (2008)
\bibitem{Hsu10}
E. N. Saridakis, JCAP 0804, 020 (2008)
\bibitem{Hsu11}
E. N. Saridakis, Phys. Lett. B 661, 335 (2008)

\bibitem{Cai1}
R.G. Cai, Phys. Lett. B 657, 228 (2007)
\bibitem{Cai12}
H. Wei and R.G. Cai, Phys. Lett. B 660, 113 (2008)
\bibitem{Cai13}
H. Wei and R.G. Cai, Eur. Phys. J. C 59, 99 (2009)

\bibitem{Ghost1}
F.R. Urban, A.R. Zhitnitsky, Phys. Rev. D 80, 063001 (2009)
\bibitem{Ghost12}
F.R. Urban, A.R. Zhitnitsky, JCAP 09, 018 (2009)
\bibitem{Ghost13}
F.R. Urban, A.R. Zhitnitsky, Phys. Lett. B 688, 9 (2010)
\bibitem{Ghost14}
F.R. Urban, A.R. Zhitnitsky, Nucl. Phys. B 835, 135 (2010)
\bibitem{Ghost15}
N. Ohta, Phys. Lett. B 695, 41 (2011)
\bibitem{Ghost16}
R.G. Cai, Z.L. Tuo, H.B. Zhang, arXiv:1011.3212

\bibitem{Ghost2}
A. Sheykhi, A. Bagheri, Europhys. Lett. 95, 39001 (2011)
\bibitem{Ghost21}
E. Ebrahimi, A. Sheykhi, Phys. Lett. B 705, 19 (2011)
\bibitem{Ghost22}
E. Ebrahimi, A. Sheykhi, Int. J. Mod. Phys. D 20, 2369 (2011)
\bibitem{Ghost23}
A. Sheykhi, M. Sadegh Movahed, Gen. Relativ. Gravit. [DOI:10.1007/s10714-011-1286-3]

\bibitem{Chao-Jun}
Chao-Jun Feng, Xin-Zhou Li, Ping Xi, JHEP 1205, 046 (2012)
\bibitem{Chao-Jun1}
Chao-Jun Feng, Xin-Zhou Li, Xian-Yong Shen, Phys.Rev. D87, 023006 (2013)
\bibitem{Chao-Jun2}
 Chao-Jun Feng, Xin-Zhou Li, Xian-Yong Shen, Mod.Phys.Lett. A27, 1250182 (2012)

\bibitem{Cai}
R.G. Cai, Z.L. Tuo, Y.B. Wu, Y.Y. Zhao, Phys Rev. D 86, 023511 (2012)

\bibitem{Hao}
WEI Hao, Common. Theory. Phys. 56, 972-980 (2011)

\bibitem{Hao2}
H.Wei, Nucl. Phys. B 845, 381 (2011)

\bibitem{Martiros2}
Martiros Khurshudyan, arXiv:1302.1220, (2013)
\bibitem{Martiros3}
Martiros Khurshudyan,  arXiv:1301.4990, (2013)
\bibitem{Martiros4}
 Martiros Khurshudyan, arXiv:1301.1021, (2013)
\bibitem{Martiros5}
Martiros Khurshudyan, arXiv:1301.0005, (2013)

\bibitem{Saridakis}
Xi-ming Chen, Yungui Gong and Emmanuel n. Saridakis, arXiv:1111.6743v2, (2012)

\bibitem{Ujjal}
Arundhati Das, Surajit Chattopadhyay, Ujjal Debnath, Found Phys 42, 266-283 (2012)
\end{thebibliography}
\end{document}